\documentclass{iaus}
\usepackage{graphicx}
\renewcommand{\theequation}{\arabic{equation}}

\title[Gravitomagnetic effects of sphere with an equatorial mass current] %% give here short title %%
{Gravitomagnetic effects of a massive and slowly rotating sphere with an equatorial mass current on orbiting test particles}

\author[L. Casta{\~n}eda, F. Fandi\~no, W. Almonacid, E. Su\'arez \& G. Pinz\'on]  %% give here short author list %%
{Leonardo Casta{\~n}eda$^1$, Fernando Fandi{\~n}o$^{1,2}$, William Almonacid$^1$, Edilberto Su\'arez$^2$ \and Giovanni Pinz\'on$^1$\footnote{email: {\tt lcastanedac@unal.edu.co, jffandilloc@unal.edu.co}}}

\affiliation{$^1$Observatorio Astron{\'o}mico Nacional, Universidad Nacional de Colombia, Bogot{\'a}, Colombia \\[\affilskip]
$^2$Universidad Distrital ``Francisco Jose de Caldas'', Bogot{\'a}, Colombia} 

\pubyear{2009}
\volume{261}  %% insert here IAU Symposium No.
\pagerange{1--3}
\jname{Relativity in Fundamental Astronomy:
Dynamics, Reference Frames, and Data Analysis}
\editors{Sergei Klioner, P. Kenneth Seidelmann \& Michael Soffel, eds.}
\begin{document}

\maketitle

\begin{abstract}
Within the framework of linearized Einstein field equations we compute the gravito-magnetic effects on a test particle orbiting a slowly rotating, spherical body with a rotating matter ring fixed to the equatorial plane. Our results show that the effect on the precession of particle orbits is increased by the presence of the ring.

\keywords{Gravitation, Solar System}
\end{abstract}


Analogy between classical electrodynamics and the linearized Einstein field equations has been largely studied (\cite[Misner, Thorne \& Wheeler 1973]{Misner}). This analogy was recently revised by \cite[Franklin \& Baker (2007)]{Franklin} in the frame of a test particle orbiting a slowly rotating sphere. In this work we have introduced a massive rotating equatorial ring surrounding the sphere in order to simulate the effect of an axisymmetric mass distribution.

\begin{figure}[t]
% \vspace*{-2.0 cm}
\begin{center}
 \includegraphics[width=2.2in]{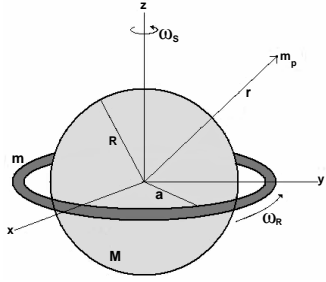} 
% \vspace*{-1.0 cm}
 \caption{Spinning sphere with radius $R$, angular frequency $\omega_S$ and equatorial mass current}
  \label{fig1}
\end{center}
\end{figure}

The gravito-electric and gravito-magnetic potentials obey a Maxwell-like set of equations (\cite[Franklin \& Baker (2007)]{Franklin}) directly derived from linearized General Relativity (\cite[Soffel 1989]{Soffel}). For a sphere with Newtonian angular momentum  $\ell_S=\frac{2}{5}MR^2\omega_S$, the corresponding potentials are given by:

\begin{equation}
\label{potS}
	V_S(r)=-\frac{GM}{r}\;, \qquad  \mathbf{A}_S(\mathbf{r})=-\frac{G}{c^2}\frac{\ell_S}{2}\frac{\sin\theta}{r^2}\hat{\varphi}\;,
\end{equation}

On the other hand, for the ring with Newtonian angular momentum $\ell_R=ma^2\omega_R$, we have for the potentials:

\begin{eqnarray}
\label{potR}
V_R(\mathbf{r})&=& -Gm \sum_{n=0}^{\infty} \frac{(-1)^n(2n)!}{2^{2n}(n!)^2} P_{2n}(\cos\theta) \frac{a^{2n}}{r^{2n+1}}\nonumber\\
	\mathbf{A}_R(\mathbf{r})&=& -\frac{G}{c^2}\ell_R \sum_{n=0}^{\infty} (-1)^n\frac{(2n)!}{(2n+2)!} \frac{(2n+1)!}{(2^{n}n!)^2} P_{2n+1}^1(\cos\theta) \frac{a^{2n}}{r^{2n+2}} \hat{\varphi}\;,
\end{eqnarray}

Restricting the discussion to the mass-monopole and spin dipole, the gravito-electric and gravito-magnetic potentials for the composite system take the form:

\begin{eqnarray}
\label{pot}
V(\mathbf{r})&=& -G \frac{(M+m)}{r}\;,\nonumber\\
	\mathbf{A}(\mathbf{r})&=& -\frac{G}{2c^2}(\ell_S+\ell_R)\frac{\sin\theta}{r^2} \hat{\varphi}\;,
\end{eqnarray}

The Lagrangian for a test particle in the field of these potentials (neglecting $c^{-2}$ terms from the Schwarzschild problem) reads:
\begin{equation}
\label{Lag}
	L(\mathbf{r},\mathbf{v},t)=\frac{m_p}{2}\left(\dot{r}^2+r^2\dot{\theta}^2+r^2\sin^2\theta\dot{\varphi}^2\right)+G m_p\frac{(M+m)}{r}
	-\frac{Gm_p}{2c^2}\frac{(\ell_S+\ell_R)}{r}\sin^2\theta \dot{\varphi}\;, 
\end{equation}

Therefore, the conserved quantities are: the $z$-component of the angular momentum and the energy of the particle. Both quantities constraint the initial conditions for the test particle.
\vspace{-0.1cm}
\begin{eqnarray}
\label{JyE}
J_z&=&\frac{\partial L}{\partial \dot{\varphi}}=m_p\left[r^2\dot{\varphi}-\frac{G}{2c^2}\frac{(\ell_S+\ell_R)}{r}\right]\sin^2\theta\;,\nonumber\\
E&=&\frac{m_p}{2}\left(\dot{r}^2+r^2\dot{\theta}^2+r^2\sin^2\theta\dot{\varphi}^2\right)-Gm_p\frac{(M+m)}{r}\;,
\end{eqnarray}

From equation (\ref{Lag}) we obtain the equations of motion for the test particle:
\vspace{-0.05cm}
\begin{eqnarray}
\label{eqmov}
\ddot{r}&=&r\dot{\theta}^2+r\sin^2\theta\dot{\varphi}^2-G\frac{(M+m)}{r^2}+\frac{G}{2c^2}\frac{(\ell_S+\ell_R)}{r^2}\sin^2\theta\dot{\varphi}\;,\nonumber\\	\ddot{\theta}&=&\frac{\sin2\theta}{2}\dot{\varphi}^2-\frac{2}{r}\dot{r}\dot{\theta}-\frac{G}{2c^2}\frac{(\ell_S+\ell_R)}{r^3}\sin2\theta\dot{\varphi}\;,\nonumber\\ 
\ddot{\varphi}&=&-\frac{2}{r}\dot{r}\dot{\varphi}-2\cot\theta\dot{\theta}\dot{\varphi}+\frac{G}{c^2}\frac{(\ell_S+\ell_R)}{r^3}\cot\theta\dot{\theta}-\frac{G}{2c^2}\frac{(\ell_S+\ell_R)}{r^4}\dot{r}\;, 
\end{eqnarray}

In order to get the trajectory of the test particle we conducted numerical integrations of equations (\ref{eqmov}). It is important to note that the term related with the matter current, originating in the gravito-magnetic field, is a perturbation of the gravito-electric field. In fact, the gravito-magnetic terms are of order $(v/c)^2$, i.e., the effect is important for large timescales. Gravito-magnetic effects onto the system are ``amplified'' by using $c=1$. We constraint our simulations to planar and closed orbits setting $\theta=\pi/2$ and $\dot{\theta}=0$. 3D simulations will be shown in a forthcoming paper (\cite[Casta\~neda et al. 2009]{Castaneda}). For the orbits in the plane, following the Bertrand's Theorem (\cite[Goldstein, Polle \& Safko 2000]{Goldstein}), we impose an initial angular velocity for the test particle to obtain elliptic orbits. Initial conditions are therefore those corresponding to the perihelion, ($r_p$), where $\varphi=0$, $\dot{r}=0$ and $\dot{\varphi}$ are obtained from the first integrals of motion given by equations (\ref{JyE}).

\begin{figure}[t]
% \vspace*{-2.0 cm}
\begin{center}
 \includegraphics[width=3.2in]{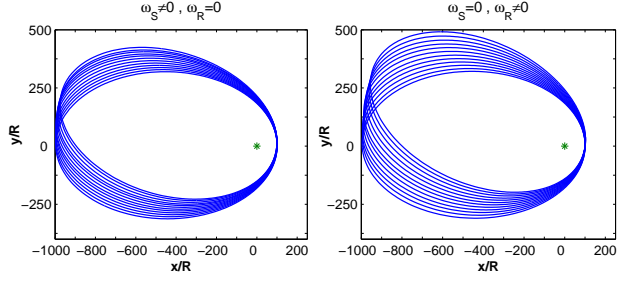} 
% \vspace*{-1.0 cm}
 \caption{Left panel: Orbits of the test particle in the x-y plane for the test particle with the ring at rest. Right panel: Sphere at rest. Parameters used in the simulation: $c=1$, $M=10^{24} kg$, $R=2*10^4 m$, $m=M/4$, $a=1.5R$, $r_p=100R$, $ra=1000R$ y $m_p=1kg$.}
   \label{fig2}
\end{center}
\end{figure}

\begin{figure}[ht]
% \vspace*{-2.0 cm}
\begin{center}
 \includegraphics[width=5in]{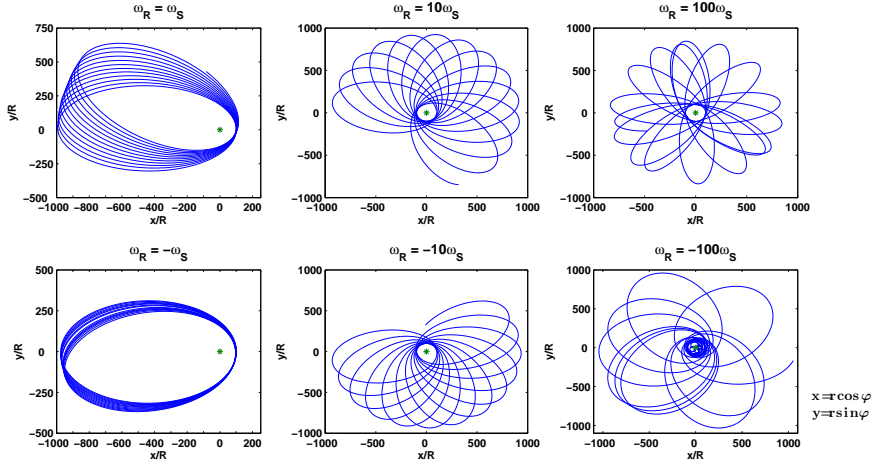} 
% \vspace*{-1.0 cm}
 \caption{Orbits of the test particle in the x-y plane for  $\omega_S=2*10^{-8} s^{-1}$ and $\omega_R=k\omega_S$ with $k=-100,-10,-1,1,10,100$.}
   \label{fig3}
\end{center}
\end{figure}

Numerical simulations were conducted using a Runge-Kutta method implemented in Matlab $7.0$ and compared with a 3D code written in C. From Figure (\ref{fig2}) the ring's gravito-magnetic contribution to the orbital precession is obvious, which is the central issue of this paper. Our main goal has been to study a system of astrophysical interest; our results are a first step to understand the main features from gravito-magnetic contributions. Gravito-magnetic effects, such as The Lense-Thirring one, follow directly from our simulations, as will be shown in \cite[Casta\~neda et al. 2009]{Castaneda}. Finally, Figure (\ref{fig3}) shows a family of trajectories for different ring velocities for a fixed sphere rotation velocity. The upper panel gives the effect of corotation between the sphere and the ring; the lower one shows the effect of counter-rotating bodies. The result is the precession of the orbit as was shown in a special case by \cite[Franklin \& Baker (2007)]{Franklin}. However, we note that the effect of the equatorial mass current is to increase the precession velocity. In fact, as the velocity of the ring increases, the effect on the precession is more noticeable. These results motivate us to extend the present problem to further studies.    

\end{document}